\documentclass[english, reprint, aps, prl, floatfix,showpacs, 10pt]{revtex4-1}
\usepackage[T1]{fontenc}
\usepackage{babel}
\usepackage{units}
\usepackage{amsmath}
\usepackage{graphicx}
\usepackage{esint}
\usepackage[T1]{fontenc}
\usepackage{lmodern}
\usepackage[latin9]{inputenc}
\usepackage{bm}

\begin{document}

\title{Effect of Dilute Random Field on Continuous-Symmetry Order Parameter}

\author{T. C. Proctor and E. M. Chudnovsky}

\affiliation{Physics Department, Lehman College, City University of New York \\
 250 Bedford Park Boulevard West, Bronx, New York 10468-1589, USA}

\date{\today}

\begin{abstract}
XY and Heisenberg spins, subjected to strong random fields acting at few points in space with concentration $c_r \ll 1$, are studied numerically on 3d lattices containing over four million sites. Glassy behavior  with strong dependence on initial conditions is found. Beginning with a random initial orientation of spins, the system evolves into ferromagnetic domains inversely proportional to $c_r$ in size. The area of the hysteresis loop, $m(H)$, scales as $c_r^2$. These findings are explained by mapping the effect of strong dilute random field onto the effect of weak continuous random field. Our theory applies directly to ferromagnets with magnetic impurities, and is conceptually relevant to strongly pinned vortex lattices in superconductors and pinned charge density waves.  
\end{abstract}


\maketitle

The random-field problem has a long history. In the early 1970s, Larkin argued \cite{Larkin-JETP1970}  that random pinning of flux lattices in superconductors leads to a finite translational correlation length which determines the critical current \cite{Blatter-RMP1994}. A more general statement was made by Imry and Ma \cite{Imry-Ma-PRL1975}, who argued that a static random field, regardless of strength, breaks the system into domains  below $d=4$ spatial dimensions. Besides superconductors, the concept of Imry-Ma domains was applied to random magnets \cite{Pelcovits,Patterson,EC-PRB86}, disordered antiferromagnets \cite{Fishman}, spin-glasses \cite{Binder}, arrays of magnetic bubbles \cite{bubbles}, charge-density waves (CDW) \cite{Efetov-77,Lee-PRB78,Lee-PRB79,Gruner-RMP88}, liquid crystals \cite{LC}, the superconductor-insulator transition \cite{SI}, and superfluid $^{3}$He-A in aerogels \cite{Volovik,Li}. In the early 1980s the validity of the Larkin-Imry-Ma argument was questioned by the renormalization group treatment \cite{Cardy-PRB1982,Villain-ZPB1984}. Scaling and replica-symmetry breaking methods \cite{Nattermann,Kierfield,Korshunov-PRB1993,Giamarchi-94,Giamarchi-95,Nattermann-2000,Feldman,LeDoussal-Wiese-PRL2006,LeDoussal-PRL2006,LeDoussal-PRL07,Bogner},
as well as the variational approach \cite{Orland-EPL,Garel-PRB}, suggested
a power-law decay of correlations at large distances, indicating a type of topological order (so called ``Bragg'' or ``elastic'' glass) which is robust against weak collective pinning. It was argued \cite{Fisher-PRL1997} that the energy associated with vortex loops prevented the $XY$ $3d$ random-field system from completely disordering. The interpretation of decoration \cite{Murray-PRL90,Murray-PRB91} and neutron diffraction \cite{Klein-Nature2001} experiments on flux lattices has been hampered by the fact that for weak disorder the correlation length can be very large, making it difficult to distinguish between large defect-free domains and a Bragg glass. Numerical work on $1d$  \cite{SL-PRB86, DC-PRB1991}), $2d$ \cite{Dieny-PRB1990,Fisher-PRL99}), and $3d$ \cite{Gingras-Huse-PRB1996,GCP-EPL13,GCP-PRB13} systems with quenched randomness established strong non-equilibrium effects, such as hysteresis and dependence on initial conditions. Recently, it has been argued and confirmed numerically on lattices containing up to one billion sites \cite{PGC-PRL2014} that long-range correlations in the presence of a weak random field are controlled by topology. For a fixed-length $n$-component order parameter in $d$ dimensions the behavior of the system is fully reversible, characterized by the exponential decay of correlations, at $n>d+1$. Meanwhile, at $n\leq d+1$ topological defects leads to metastabillity and glassy behavior that depends on the initial condition.  

In general, for spin systems it is hard to come up with a physical mechanism that would generate a random field (different from magnetic anisotropy) linearly interacting with a spin at every lattice site. An exception is anisotropic antiferromagnet with fluctuating exchange interaction in a uniform magnetic field \cite{Fishman}. However, if disorder due to, e.g., magnetic impurities is confined to few points in space, then the exchange interaction $-J_r{\bf S}\cdot {\bf s}$ of the lattice spin ${\bf s}$ with the nearby impurity spin ${\bf S}$ is equivalent to the interaction with the magnetic field ${\bf h} = J_r {\bf S}$. The model of a ferromagnet with dilute impurity spins with frozen random orientations is, therefore, equivalent to the random-field model in which the field only acts on spins located at few points in space. The XY spin model of this kind was recently studied by Okamoto and Millis \cite{Millis-CondMat}  in connection with the problem of strong pinning of the CDW by dilute pinning centers. The computed correlations extended far beyond the average distance between strong pinning centers. This study was inspired by the experimental evidence of extended phase correlations in the CDW in NbSe$_2$ \cite{Millis-PRL15} that disagreed with earlier theoretical predictions \cite{Efetov-77,Lee-PRB78,Lee-PRB79,Gruner-RMP88}. 

Our study of XY and Heisenberg spins on $3d$ lattices containing over four million sites is focused on metastability. The measure of metastability is the area of the hysteresis loop. We find that it scales as $c_r^2$, where $c_r \ll 1$ is the concentration of spins subjected to strong dilute random fields. If one starts with random initial conditions (RIC) for the spins the system evolves towards short-range order with exponential decay of correlations and with the correlation length inversely proportional to $c_r$. In accordance with the findings of Okamoto and Millis, as well as with experiments on NbSe$_2$, the correlation length that we obtained is much greater than the average distance, $1/c_r^{1/3}$, between the impurities. We provide a simple physical explanation to these findings. 

We consider the spin Hamiltonian 
\begin{equation}
{\cal H}=-\frac{1}{2}\sum_{ij}J_{ij}{\bf s}_{i}\cdot{\bf s}_{j}-\sum_{i,r}{\bf h}_{i}\cdot{\bf s}_{i}\delta_{ir}-{\bf H}\cdot\sum_{i}{\bf s}_{i},\label{Ham-descrete}
\end{equation}
with the nearest-neighbor exchange interaction, strong dilute random field ${\bf h}$ acting at $r$-sites with concentration $c_r \ll 1$, and the external field ${\bf H}$, on lattices of a few millions spins ${\bf s}_{i}$ of length $s$.  In numerical work we consider infinite ${\bf h}$, which implies that some spins with concentration $c_r$ are frozen in random directions. Our numerical method combines sequential rotations of spins towards the direction of the local effective field, ${\bf H}_{i,{\rm eff}}=\sum_{j}J_{ij}{\bf s}_{j}+\sum_{r}{\bf h}_{i}\delta_{ir}+{\bf H}$,
with energy-conserving spin flips: ${\bf s}_{i}\rightarrow2({\bf s}_{i}\cdot{\bf H}_{i,{\rm eff}}){\bf H}_{i,{\rm eff}}/H_{i,{\rm eff}}^{2}-{\bf s}_{i}$, applied with probabilities $\alpha$ and
$1-\alpha$ respectively; $\alpha$ playing the role of the relaxation
constant. This method is physically equivalent to slow cooling. Its fast convergence to the final state that is no longer relaxing has been demonstrated in Refs. \onlinecite{GCP-EPL13,GCP-PRB13,PGC-PRL2014}.  

\begin{figure}
\begin{centering}
\includegraphics[width=7.5cm]{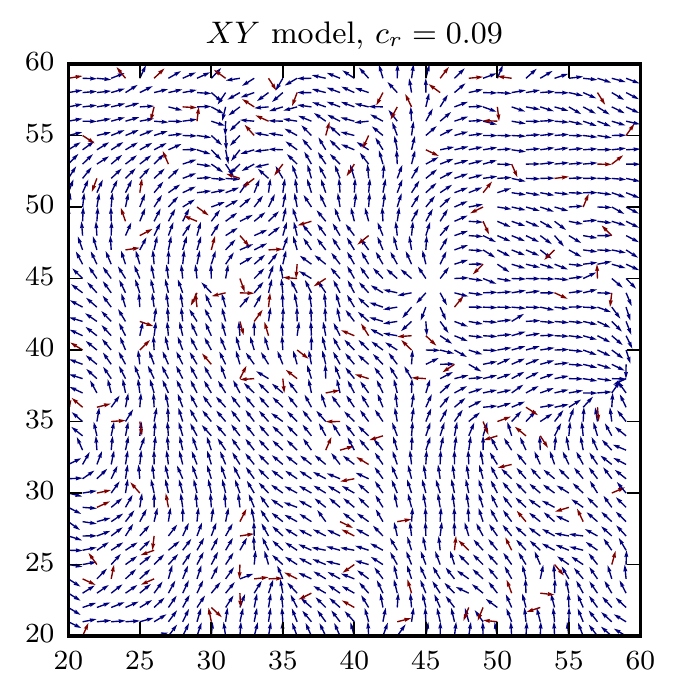}
\par\end{centering}
\caption{Color online: Snapshot of XY spins in one layer of a $3d$ lattice after relaxation from random initial orientation in the presence of dilute random field. Spins whose orientations are frozen by the random field are shown in red.}
\label{snapshot} 
\end{figure}
\begin{figure}
\begin{centering}
\includegraphics[width=7.5cm]{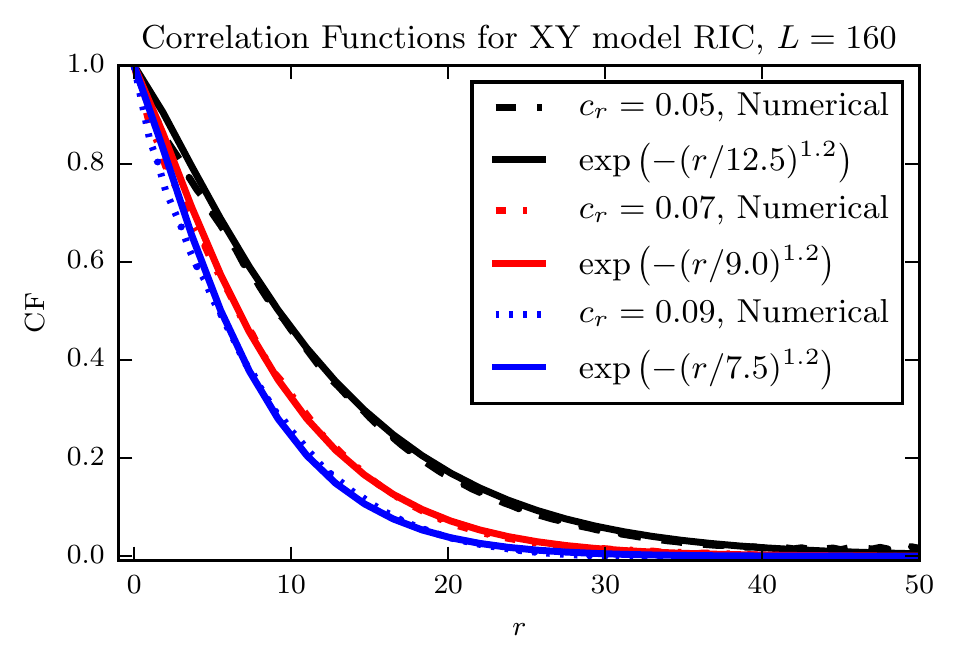}{a\label{CFXY}}
\includegraphics[width=7.5cm]{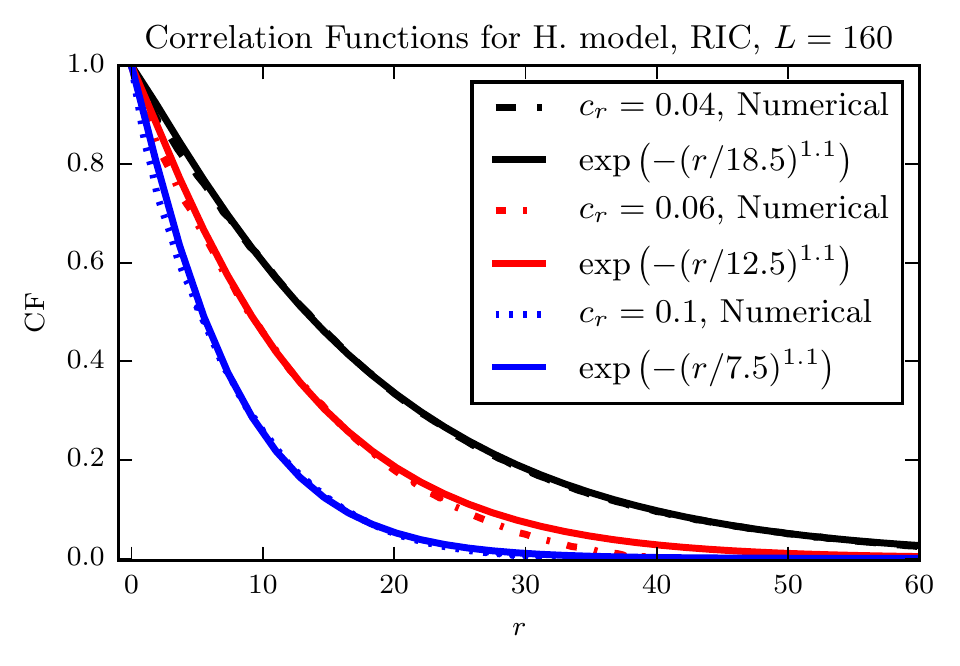}{b\label{CFHeis}}
\par\end{centering}
\caption{Color online: Spin-spin correlation function (dashed lines) after relaxation from random initial orientation of spins in the presence of dilute random field. Solid lines provide the corresponding fit by the exponential. Distances are given in lattice units, with $L$ being the size of the $L\times L \times L$ system. a) $3d$ $XY$ spin model. b) $3d$ Heisenberg  spin model.}
\label{CF} 
\end{figure}
\begin{figure}
\begin{centering}
\includegraphics{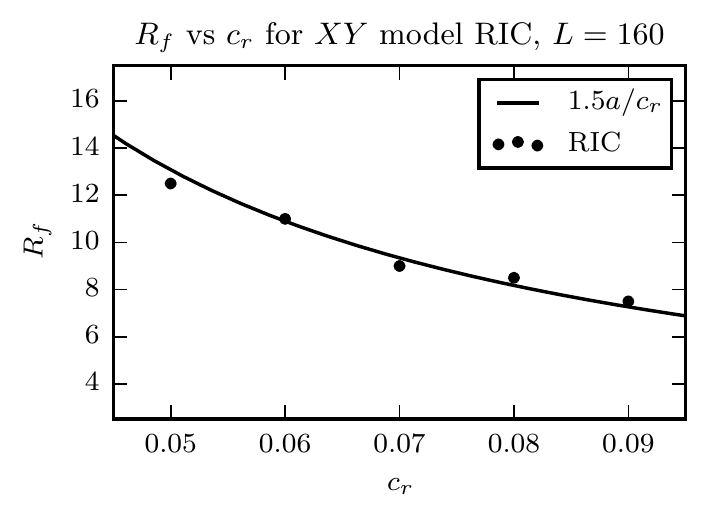}{a\label{RfXY}}
\includegraphics{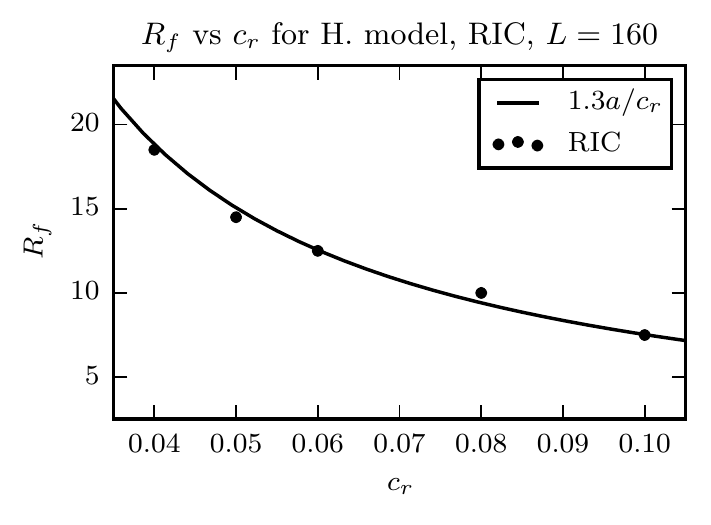}{b\label{RFHeis}}
\par\end{centering}
\caption{Dependence of the ferromagnetic correlation length on the concentration of random field sites for the state obtained by evolution from random initial orientation of spins. a) $3d$ $XY$ spin model. b) $3d$ Heisenberg  spin model.}
\label{Rf} 
\end{figure}

We begin with evolution of the system from random initial orientation of spins. The randomly chosen $c_r$ fraction  of spins remain frozen in random directions, while other spins are allowed to rotate and relax to some final state in which the total magnetization is no longer changing. A snapshot of such a state is shown in Fig. \ref{snapshot}. Presence of topological defects and short-range order is apparent. The corresponding spin-spin correlation function and its fit by the exponential are shown in Fig. \ref{CF}.
This fit allows one to extract the ferromagnetic correlation length $R_f$. Its dependence on the concentration of impurities, $c_r$, is shown in Fig. \ref{Rf}.
The $1/c_r$ dependence of $R_f$ provides a good fit to the numerical data for both the $XY$ model with two spin components and the Heisenberg model with three spin components.

\begin{figure}
\begin{centering}
\includegraphics[width=7.5cm]{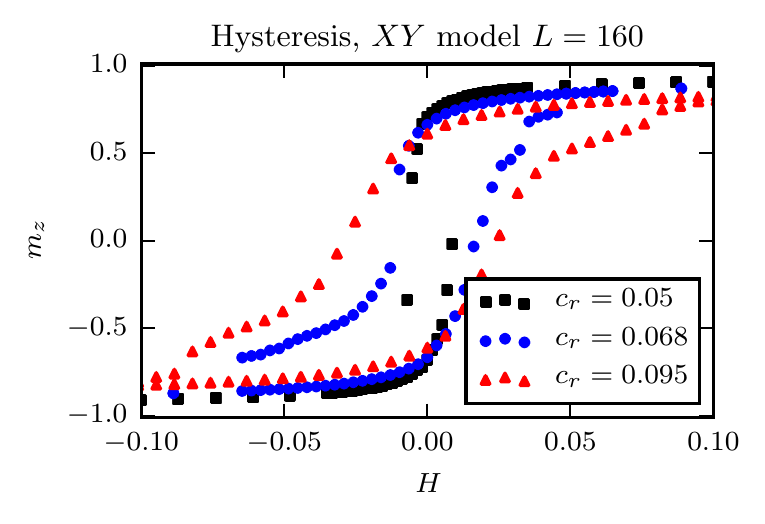}{a}
\includegraphics[width=7.5cm]{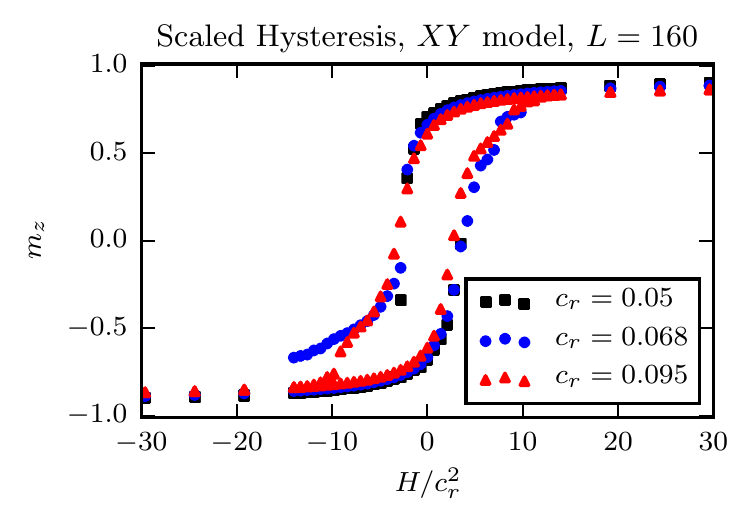}{b}
\par\end{centering}
\caption{Color online: Hysteresis loops for the $3d$ $XY$ model with dilute strong random field for different concentrations of the random-field sites, $c_r$. a) Unscaled per-spin magnetization $m_z$ vs $H$. b) Scaled per-spin magnetization $m_z$ vs $H/c_r^2$.}
\label{hysteresis-Heis} 
\end{figure}
\begin{figure}
\begin{centering}
\includegraphics[width=7.5cm]{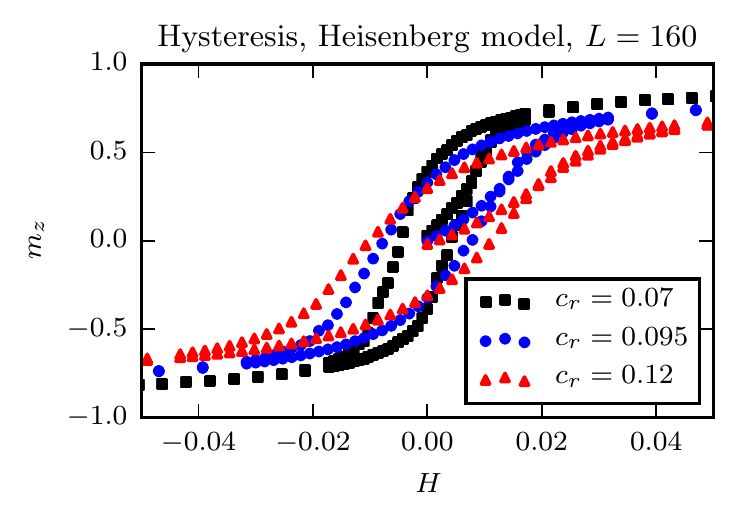}{a}
\includegraphics[width=7.5cm]{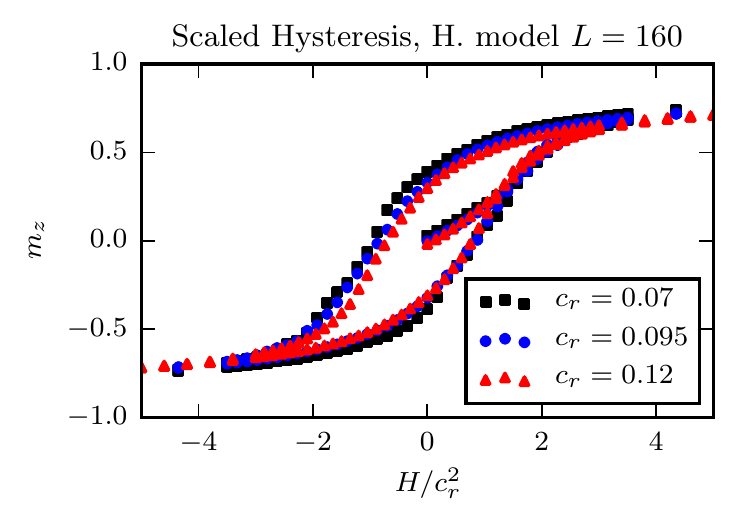}{b}
\par\end{centering}
\caption{Color online: Hysteresis loops for the $3d$ Heisenberg model with dilute strong random field for different concentration of the random-field sites, $c_r$. a) Unscaled per-spin magnetization $m_z$ vs $H$. b) Scaled per-spin magnetization $m_z$ vs $H/c_r^2$. }
\label{hysteresis-XY} 
\end{figure}
The state obtained by the relaxation from collinear initial conditions (with all spins initially oriented in one direction) always has non-zero magnetization. This is an indication of metastability and glassy behavior caused by the dilute random field. There may be as little profit in looking for the ``needle-in-the-haystack'' ground state of such a system as in looking for the ground-state configuration of domains of a conventional permanent magnet with a macroscopic number of defects that pin domain walls. We, therefore, turn to the magnetic hysteresis as a measure of metastability. We compute the magnetization per spin $m_z$ induced by the field applied in the $Z$-direction. Hysteresis loops for $XY$ and Heisenberg models are shown in Figs. \ref{hysteresis-XY} and \ref{hysteresis-Heis}.
The scaling in terms of $H/c_r^2$ is very good, implying that the coercive field and the area of the hysteresis loop roughly scale as $c_r^2$. 

Our findings can be explained in a simple manner by employing the Larkin-Imry-Ma argument. It is convenient to rescale the problem to subvolumes of size $r_h \sim 1/c_r^{1/3}$, which  represents the average distance between the random-field sites. In the lattice units, the effective random field per spin inside such a volume is of order $h_{r_h} \sim h/r_h^3$. If we consider a group of spins ferromagnetically correlated within a volume of size $R_f \gg r_h$, the exchange energy per spin inside this volume is of order $J/R_f^2$, while statistical fluctuation of the Zeeman energy per spin is of order $-h_{r_h}/(R_f/r_h)^{3/2}$. Minimization of the total energy with respect to $R_f$ gives $R_f \sim r_h^3(J/h)^2 \propto 1/c_r$, in accordance with our numerical results. In a similar manner one can compute the scaling of the coercive field. It roughly equals the field needed to overcome the effective field $h_{R_f} \sim h_{r_h}/(R_f/r_h)^{3/2}$ which aligns the spins on the scale $R_f$. Substituting  $h_{r_h}$ and $R_f$ from above, one obtains $h_{R_f} \propto 1/r_h^6 \sim c_r^2$. This scaling of the coercive field and, consequently, of the area of the hysteresis loop is in agreement with our numerical results.

Our model directly applies to a ferromagnet with magnetic impurities having strong single-ion anisotropy that freezes their spins in random directions. We assume that such impurities interact strongly via ferromagnetic or antiferromagnetic exchange with the surrounding spins. Note that every magnetic system also has relativistic interactions, such as dipole-dipole and magneto-crystalline anisotropy terms in the Hamiltonian, that would break the system into domains even in the absence of the spin impurities generating strong random fields. However, for sufficiently large concentration of magnetic impurities the formation of the short-range order would be dominated by their strong exchange interaction with the surrounding spins rather than by the weak relativistic interactions in the system. In general, magnetic impurities must be of practical significance if the short-range-order ferromagnetic correlation length, $R_f \propto 1/c_r$, that they are responsible for, is smaller than the would-be average domain size without the magnetic impurities. 

While the spin problem does not map exactly onto the flux-lattice problem or the CDW problem (see discussion of these issues in, e.g., Refs. \onlinecite{Gingras-Huse-PRB1996} and \onlinecite{Millis-CondMat}), all three problems are conceptually similar. Consequently, our study supports statements of Refs. \onlinecite{Millis-PRL15} and \onlinecite{Millis-CondMat} that the correlation length in the CDW problem with diluted strong pinning centers is large compared to the average distance between the centers. Our prediction for the correlation length is $R_f \sim 1/c_r \gg r_h \sim 1/c_r^{1/3}$. It would be interesting to test this prediction in NbSe$_2$. It can also be tested on flux lattices that are pinned by strong pinning centers with concentration $c_r \ll 1$. Scaling of the critical current as $j_c \sim 1/R_f^2$ \cite{Blatter-RMP1994} would then imply $j_c \propto c_r^2$. 

This work was supported by the grant No. DE-FG02-93ER45487 funded by the U.S. Department of Energy, Office of Science.


\begin{thebibliography}{10}
\bibitem{Larkin-JETP1970} A. I. Larkin, Sov. Phys. JETP \textbf{31},
784 (1970).

\bibitem{Blatter-RMP1994} G. Blatter, M. V. Feigel'man, V. B. Geshkenbein,
A.I. Larkin, and V. M. Vinokur, Rev. Mod. Phys. \textbf{66}, 1125
(1994).

\bibitem{Imry-Ma-PRL1975} Y. Imry and S.-k. Ma, Phys. Rev. Lett.
\textbf{35}, 1399 (1975).

\bibitem{Pelcovits} R. Pelcovits, E. Pytte, and J. Rudnick, Phys.
Rev. Lett. \textbf{40}, 476 (1978)

\bibitem{Patterson} J. D. Patterson, G. R. Grusalski, and D. J. Sellmyer,
Phys. Rev. B \textbf{18}, 1377 (1978).

\bibitem{EC-PRB86} E. M. Chudnovsky, W. M. Saslow and
R. A. Serota, Phys. Rev. B \textbf{33}, 251 (1986), and references
therein.

\bibitem{Fishman} S. Fishman and A. Aharony, J. Phys. C: Solid State
Phys. \textbf{12}, L729 (1979).

\bibitem{Binder} K. Binder and A. P. Young, Rev. Mod. Phys. \textbf{58},
801 (1986).

\bibitem{bubbles} R. Seshadri and R.M. Westervelt, Phys. Rev. B \textbf{46},
5142 (1992); ibid. \textbf{46}, 5150 (1992).

\bibitem{Efetov-77} K. B. Efetov and A. I. Larkin, Sov. Phys. JETP
\textbf{72}, 2350 (1977).

\bibitem{Lee-PRB78}
H. Fukuyama and P. A. Lee, Phys. Rev. B {\bf 17}, 535 (1978).

\bibitem{Lee-PRB79}
P. A. Lee and T. M. Rice, Phys. Rev. B {\bf 19}, 3970 (1979).

\bibitem{Gruner-RMP88}
G. Gr\"{u}ner, Rev. Mod. Phys. {\bf 60}, 1129 (1988). 

\bibitem{LC} T. Bellini, N. A. Clark, V. Degiorgio, F. Mantegazza,
and G. Natale, Phys. Rev. E \textbf{57}, 2996 (1998).

\bibitem{SI} E. M. Chudnovsky, Phys. Rev. Lett. \textbf{103}, 137001
(2009).

\bibitem{Volovik} G. E. Volovik, J. Low Temp. Phys. \textbf{150},
453 (2008).

\bibitem{Li} J. I. A. Li, J. Pollanen, A. M. Zimmerman, C. A. Collett,
W. J. Gannon, and W. P. Halperin, Nat. Phys. \textbf{9}, 775 (2013).

\bibitem{Cardy-PRB1982} J. L. Cardy and S. Ostlund, Phys. Rev. B
\textbf{25}, 6899 (1982).

\bibitem{Villain-ZPB1984} J. Villain and J. F. Fernandez, Z. Phys.
B - Condens. Matter \textbf{54}, 139 (1984).

\bibitem{Nattermann} T. Nattermann, Phys. Rev. Lett. \textbf{64},
2454 (1990).

\bibitem{Kierfield} J. Kierfield, T. Nattermann, and T. Hwa, Phys.
Rev. B \textbf{55}, 626 (1997).

\bibitem{Korshunov-PRB1993} S. E. Korshunov, Phys. Rev. B \textbf{48},
3969 (1993).

\bibitem{Giamarchi-94} T. Giamarchi and P. Le Doussal, Phys. Rev.
Lett. \textbf{72}, 1530 (1994).

\bibitem{Giamarchi-95} T. Giamarchi and P. Le Doussal, Phys. Rev.
B \textbf{52}, 1242 (1995).

\bibitem{Nattermann-2000}  T. Nattermann and S. Scheidl, Adv. Phys. {\bf 49}, 607 (2000).

\bibitem{Feldman} D. E. Feldman, Phys. Rev. B \textbf{61}, 382 (2000).

\bibitem{LeDoussal-Wiese-PRL2006} P. Le Doussal amd K. J. Wiese,
Phys. Rev. Lett. \textbf{96}, 197202 (2006).

\bibitem{LeDoussal-PRL2006} P. Le Doussal, Phys. Rev. Lett. \textbf{96},
235702 (2006).

\bibitem{LeDoussal-PRL07} A. A. Middleton, P. Le Doussal, and K.
J. Wiese, Phys. Rev. Lett. \textbf{98}, 155701 (2007).

\bibitem{Bogner} S. Bogner, T. Emig, A. Taha, and C. Zeng, Phys.
Rev. B \textbf{69}, 104420 (2004).

\bibitem{Orland-EPL} H. Orland and Y. Shapir, Europhys. Lett. \textbf{30},
203 (1995).

\bibitem{Garel-PRB} T. Garel, G. Lori, and H. Orland, Phys. Rev.
B \textbf{53}, R2941 (1996).

\bibitem{Fisher-PRL1997} D. Fisher, Phys. Rev. Lett. \textbf{78},
1964 (1997).

\bibitem{Murray-PRL90}
C. A. Murray, P. L. Gammel, D. J. Bishop, D. B. Mitzi, and A. Kapitulnik, Phys. Rev. Lett. {\bf 64}, 2312 (1990).

\bibitem{Murray-PRB91} 
D. G. Grier, C. A. Murray, C. A. Bolle, P. L. Gammel, D. J. Bishop, D. B. Mitzi, and A. Kapitulnik, Phys. Rev. Lett. {\bf 66}, 2270 (1991).

\bibitem{Klein-Nature2001} T. Klein, I. Joumard, S. Blanchard, J.
Marcus, R. Cubitt, T. Giamarchi, and P. Le Doussal, Nature \textbf{413},
404 (2001).

\bibitem{SL-PRB86}
R. A. Serota and P. A. Lee, Phys. Rev. B {\bf 34}, 1806 (1986). 

\bibitem{DC-PRB1991} R. Dickman and E. M. Chudnovsky, Phys. Rev.
B \textbf{44}, 4397 (1991).

\bibitem{Dieny-PRB1990} B. Dieny and B. Barbara, Phys. Rev. B \textbf{41},
11549 (1990).

\bibitem{Fisher-PRL99} C. Zeng, P. L. Leath, and D. S. Fisher, Phys.
Rev. Lett. \textbf{82}, 1935 (1999).

\bibitem{Gingras-Huse-PRB1996} M. J. P. Gingras and D. A. Huse, Phys.
Rev. B \textbf{53}, 15193 (1996).

\bibitem{GCP-EPL13} D. A. Garanin, E. M. Chudnovsky, and T. Proctor,
Europhys. Lett. \textbf{103}, 67009 (2013).

\bibitem{GCP-PRB13} D. A. Garanin, E. M. Chudnovsky, and T. Proctor,
Phys. Rev. B \textbf{88}, 224418 (2013).

\bibitem{PGC-PRL2014} T. C. Proctor, D. A. Garanin, and E. M. Chudnovsky,
Phys. Rev. Lett. \textbf{112}, 097201 (2014).

\bibitem{Millis-CondMat}
J.-I. Okamoto and A. J. Millis, cond-mat arXiv:1412.3965 (12 December 2014).

\bibitem{Millis-PRL15}
J.-I. Okamoto, C. J. Arguello, E. P. Rosenthal, A. N. Pasupathy, and A. J. Millis, Phys. Rev. Lett. {\bf 114}, 026802 (2015). 


\end{thebibliography}
\end{document}